# Compression Performance Analysis of Different File Formats

Han Yang[1],Guangjun Qin[*],Yongqing Hu[1]

[1](Smart City College, Beijing Union University, Beijing 100101, China)
[*](Corresponding author(s). Email(s):zhtguangjun@buu.edu.cn)

Han Yang：Proposing research ideas, designing research protocols, conducting experiments and drafting papers.
Yongqing Hu：Collecting, cleaning and analysing data.
Guangjun Qin：Final version of thesis revised.

## Abstract

In data storage and transmission, file compression is a common technique for reducing the volume of data, reducing data storage space and transmission time and bandwidth. However, there are significant differences in the compression performance of different types of file formats, and the benefits vary. In this paper, 22 file formats with approximately 178GB of data were collected and the Zlib algorithm was used for compression experiments to compare performance in order to investigate the compression gains of different file types. The experimental results show that some file types are poorly compressed, with almost constant file size and long compression time, resulting in lower gains; some other file types are significantly reduced in file size and compression time after compression, which can effectively reduce the data volume. Based on the above experimental results, this paper will then selectively reduce the data volume by compression in data storage and transmission for the file types in order to obtain the maximum compression yield.

# 1. Introduction

With the widespread use of computer technology in various fields, a large amount of data has been generated that needs to be stored, computed and transmitted, and the scale of data has been exploding year by year, indicating that it has entered the era of massive data [1]. These massive amounts of data need to be quickly migrated to computing and storage devices, leading to an increasingly acute conflict between data transmission and business needs [6], with major challenges from bandwidth requirements to transmission integrity [5].

One effective way to improve the performance of massive data transfers is to reduce data size, i.e. reduce network load and transmission latency by compressing data for transmission. By compressing the data size and converting it into a more compact format, the amount of data effectively transmitted is reduced, while also reducing the time and storage space required for transmission [4], thereby reducing transmission latency and cost and increasing the speed of data transmission [2]. However, the compressibility of different file formats varies greatly, with some file formats having a highly compressible structure, such as repetitive characters in text files that can be exploited for better compression. In contrast, image files often have similar adjacent pixels and redundancy, but recompression of already compressed image files has limited effect. Audio files are often compressed using lossy compression algorithms, so further compression is less effective. Therefore, in order to investigate the compression gains of different file formats for data transmission, the performance of data storage and transmission is improved. In this paper, compression experiments are conducted on 22 different file formats to study the compression gains of different file formats and to provide a reference basis for massive data transmission.

The contributions of this paper are as follows:

1. a collection of 22 file formats, including MP4, MP3, BMP, HDF5, etc., totalling 178 GB;

2. the above datasets were investigated using the zlib compression algorithm;

3. the experiments revealed that some file types, such as audio, video and images, are relatively poorly compressed and take longer to compress, resulting in relatively low gains from compression. This is due to the fact that these file types usually have high-dimensional data structures and complex information content, making it difficult to achieve high compression ratios during the compression process. However, not all file types face the same problem. For some other file types, such as text and documents, the file size is significantly reduced after compression, and the compression time is shorter, allowing for an effective reduction in data volume.

The subsequent sections of this paper are structured as follows:

1. section2 describes the related work on file compression;

2. section3 describes the experimental dataset file sources, compression performance metrics and experimental methodology;

3. section4 describes the experimental environment and tools;

4. section5 presents a graphical analysis of the experimental results;

5. section6 provides a summary and outlook.

# 2. Literature review

File compression is an important data processing technique that improves storage efficiency and transfer speed by reducing the storage size of files and reducing disk space usage. Compression algorithms can be divided into lossless compression, which completely restores the original file, and lossy compression, which loses a certain amount of information to obtain a higher compression ratio. Lossy compression is often used for multimedia information, and lossless compression algorithms are used for text files that require integrity. gzip [17] is a widely used compression program that can be used to compress large, lesser-used files to save disk space, with a compression ratio of around 3 to 10 times, which can significantly reduce network bandwidth consumption by servers. bzip2 [18] uses the Burrows- Wheeler block sorting text compression algorithm and Huffman encoding method, the compression ratio is usually better than LZ77/LZ78 based compression software, which can compress files to within 10% to 15%. lzma [7] is a modified and optimised Deflate and LZ77 algorithm, using a dictionary encoding mechanism similar to LZ77, which has a higher compression ratio than bzip2. The Zlib library [19] provides high compression ratios and lossless compression using the DEFLATE algorithm based on a sliding window mechanism to process data in bytes and achieve compression by replacing strings, and is widely used in several fields such as Chinese retrieval, data communication and data acquisition [3].

In applications, Li Ming et al. used a lossless compressed transmission algorithm based on a sparse representation of the signal to improve the amount of uploaded information per unit of time [40]. Wang Julong et al. used the Steim2 compression algorithm and FTP communication protocol to achieve real-time data compression and transmission, which significantly improved the data transmission efficiency [41]. Yang Jingfeng et al. proposed a data compression method based on improved Huffman coding technology to achieve data compression, transmission, parsing and decompression [43]. Peng Chong et al. proposed a compression scheme based on node similarity clustering (SSCDCT), which reduced data transmission and energy consumption and extended network lifetime by aggregating similar nodes and compression algorithms [44]. Ma Xingming et al. proposed a compression and storage method for massive multivariate heterogeneous smart grid data based on state estimation, which solved the problems of large compression errors and long running time [45]. Wang He et al. proposed a power quality data compression and storage method based on distributed compression sensing and edge computing, which solved the problem of difficult division of power quality data and harmonic pollution, and achieved high precision compression and storage space saving [46].

It can be seen that compression technology has significant benefits in the field of mass data storage and transmission. However, the compression benefits have not been systematically studied and evaluated for massive files of different formats. Therefore, the conduct of this experiment is of great importance. The research objective of this paper is to investigate the benefits of decompression for different formats of massive files in order to optimise the subsequent storage and transmission process. Through the results of this paper, valuable insights and guidance will be provided to the field of data management and transfer technology.

# 3. Methodology

## 3.1 File dataset selection and preparation

Several datasets were used to form this experiment in different formats, with the following data sources:

1. video format MP4, from the Kinetics dataset on the KAGGLE website.

2. Video formats AVI, MKV, WEBM, augmented from the Kinetics dataset by Format Factory[1]

3. Audio format MP3, from the UCI website dataset FMA: A Dataset For Music Analysis Data Set.

4. Audio formats FLAC, WAV, WMA, converted from the dataset FMA: A Dataset For Music Analysis Data Set via Format Factory.

5. Image format BMP, from the KAGGLE website Alphabet+Numbers.

6. Image format GIF from Synthea Dataset Jsons - HER on the KAGGLE website.

7. Image format JPG, from the dataset A Large-Scale CarDataset for Fine-Grained Categorization and Verification of the CVPR2015 paper.

8. Image format PNG, from the KAGGLE website dataset RSNA BreastCancer Detection - 512×512 pngs

9. Image format TIF, converted from RSNA BreastCancer Detection - 512 × 512 pngs via Format Factory.

10. Document formats DOCX, XLS, XML datasets as self-captured.

11. Document format PDF, converted from self-collected DOCX dataset via Format Factory.

12. Document format TXT, from the KAGGLE website dataset Text Classification on Emails.

13. Document format JSON, from the KAGGLE website dataset Various Pokemon Image Dataset.

The experimental dataset files are shown in Table 1:

Table 1 Experimental data set file

| Dataset name | File formats | File type |
| --- | --- | --- |
| Transferred from Kinetics dataset | Video | AVI |

---

[1] [Format Factory - a free and versatile multimedia file conversion tool (formatfactory.org)].

| | | |
|---|---|---|
| Transferred from Kinetics dataset | Video | MKV |
| Kinetics dataset[10] | Video | MP4 |
| Transferred from Kinetics dataset | Video | WEBM |
| Transferred from FMA | Audio | FLAC |
| FMA: A Dataset For Music Analysis Data Set[11] | Audio | MP3 |
| Transferred from FMA | Audio | WAV |
| Transferred from FMA | Audio | WMA |
| Metal Surface Defects Dataset[15] | Image | BMP |
| Synthea Dataset Jsons - EHR[12] | Image | GIF |
| A Large-Scale Car Dataset for Fine-Grained Categorization and Verification. (CVPR)[14] | Image | JPG |
| RSNA Breast Cancer Detection - 512x512 pngs[9] | Image | PNG |
| Transferred from A Large-Scale Car Dataset for Fine | Image | TIF |
| Transferred from DOCX | Documentation | PDF |
| Self-collection | Documentation | DOCX |
| Self-collection | Documentation | XLS |
| Self-collection | Text | XML |
| Text Classification on Emails[13] | Text | TXT |
| Various Pokemon Image Dataset[8] | | JSON |
| Self-collection | Binary | BIN |

In the experiments, the compression effect of different file formats was evaluated using self-picked and publicly available datasets totalling 91GB. These publicly available datasets cover 20 file types, including text files, image files, audio files and video files.

## 3.2 Compression performance evaluation metrics

Let the file size before compression be $\alpha$, the file size after compression be $\beta$, the compression ratio be K, the compression time be T and the decompression time be T', then the compression ratio index can be expressed as

$$K = (1 - \beta/\alpha) * 100\%$$

• Compression ratio: indicates the percentage of data reduced by compression, calculates the average compression ratio for each file type, and compares the difference in compression ratio between file types. higher values of K indicate better compression.

• Compression time: indicates the time in milliseconds it takes to compress the original file into a compressed file. Calculates the average compression time for each file type and compares the difference in compression time between file types. a smaller value of T indicates faster compression.

• Decompression time: indicates the time in milliseconds it takes for a compressed file to be decompressed into its original file. Calculate the average decompression time for each file type and compare the difference in decompression time between file types. smaller values of T' indicate faster decompression.

## 3.3 Compression methods

When choosing a suitable file compression algorithm, comparisons need to be made based on requirements and scenarios. the Zlib library, with its high compression ratio and lossless compression, achieves maximum file size reduction with the DEFLATE algorithm. Zlib's extensive support and cross-platform nature makes it easy to call its functions and interfaces for file compression and decompression operations, which makes it ideal for conducting experiments to obtain more accurate and comprehensive results. Therefore, the Zlib library was chosen for this experiment and the parameters were defined as follows: the file before compression is X and the file after compression is Y.

**The compression process can be expressed as follows: Y=Zlib(X)**

**The decompression process can be expressed as follows: X=Zlib(Y)**

The experimental steps are as follows:

1. import the test dataset;

2. invoke the Zlib library to traverse the list of files in the specified directory, perform the compression process Y=Zlib(X) and the decompression process X=Zlib(Y) on each file sample, and record performance indicators such as compression ratio K, compression time T, decompression time T', information such as file pre-compression size α and post-compression size β;

3. delete the compressed and decompressed files generated during the process;

4. writing the resulting records to a CSV file;

5. calculate the average value of each metric in the CSV file to improve the reliability of the results.

## 4. Experiment

In the experiments, the 20 file datasets were first compressed and decompressed, and the results were analysed in detail. To ensure the accuracy of the experimental results, the generated files were deleted immediately after compression and decompression to avoid taking up additional storage space. The compression performance metrics from Section 3.2 were used as evaluation criteria for this experiment.

### 4.1 Experimental environment and tools

This experiment was conducted on a 64-bit Ubuntu 22.04.2LTS computer, configured with 32.0GIB of RAM and 1T of disk capacity, and a g++-11 compiler.

### 4.2 Experimental analysis

#### 4.2.1 Full volume data sets

The Zlib library was used for compression and decompression operations for all 20 different types of file datasets, and the compression time decompression time, pre-compression file size, post-compression file size, and compression ratio were recorded for each file type. The dataset used for the experiments was approximately 91GB of data, with video files (AVI, MKV, MP4, WEBM) totalling approximately 54.4GB, audio files (FLAC, MP3, WAV, WMA) totalling approximately 12.58GB, image files (BMP, GIF, JPG, PNG, TIF) totalling approximately 9.65GB, and document files (PDF, DOCX, XLS) about a total of 0.96GB, text files (XML, TXT) about a total of 0.18GB, JSON files about 0.85GB, BIN files about 0.04GB.

The experimental results are shown in Table 2.

Table 2 Experimental data results

| Param  Type | Number | Pre-compression size (MB) | Compressed size (MB) | Compression time (MS) | Decompression time (MS) | Compression rate (%) |
|------|------|------|------|------|------|------|
| AVI | 12748 | 0.522941 | 0.508537 | 17.76 | 3.28 | 3.49% |
| MKV | 10350 | 2.105637 | 2.102279 | 54.733 | 6.656 | 0.71% |
| MP4 | 10214 | 1.469854 | 1.463328 | 38.026 | 3.997 | 0.00 |
| WEBM | 6653 | 1.200335 | 1.20309 | 33.938 | 2.162 | 0.12% |
| FLAC | 8733 | 0.581497 | 0.574143 | 13.019 | 0.963 | 1.93% |
| MP3 | 8732 | 0.13802 | 0.134574 | 4.877 | 0.972 | 2.98% |
| WAV | 8733 | 0.774194 | 0.660883 | 30.932 | 5.777 | 14.02% |
| WMA | 8733 | 0.242912 | 0.109045 | 5.96 | 1.326 | 54.61% |
| BMP | 15557 | 0.750051 | 0.132008 | 22.156 | 3.6 | 82.40% |
| GIF | 7261 | 0.08965 | 0.086222 | 3.781 | 0.678 | 4.16% |
| JPG | 10547 | 0.089225 | 0.088777 | 2.693 | 0.392 | 0.66% |
| PNG | 54707 | 0.065476 | 0.065365 | 1.996 | 0.287 | 0.28% |
| TIF | 8895 | 0.140895 | 0.135055 | 4.835 | 0.981 | 5.00% |
| PDF | 2485 | 0.02381 | 0.021898 | 1.193 | 0.225 | 8.08% |
| DOCX | 2485 | 0.051761 | 0.043766 | 2.193 | 0.389 | 14.75% |
| XLS | 15121 | 0.052615 | 0.012066 | 2.582 | 0.324 | 77.05% |
| XML | 3912 | 0.043191 | 0.003706 | 1.129 | 0.19 | 89.51% |
| TXT | 7760 | 0.002006 | 0.001005 | 0.302 | 0.67 | 42.28% |
| JSON | 10628 | 0.080233 | 0.019754 | 1.926 | 0.399 | 68.64% |
| BIN | 1500 | 0.029862 | 0.003754 | 1.861 | 0.198 | 86.93% |

A comparison of file compression ratios, compression times, and decompression times for the 20 formats is shown in Figure 1:

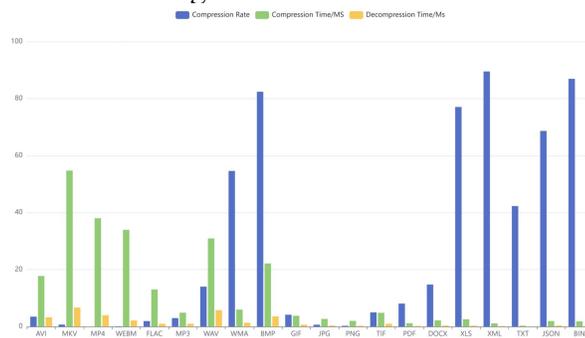

Figure 1 Comparison of compression ratio, compression time and decompression time for 20 file formats

The experimental results show that the compression rates of different file types show significant differences, with video formats (AVI, MKV, MP4, WEBM), audio formats (FLAC, MP3) and image formats (GIF, JPG, PNG, TIF) showing almost no significant change in file size after compression, with lower gains; in contrast, other file types are significantly reduced after compression, and The compression time is also shorter, which can effectively reduce the amount of data. The specific experimental analysis is as follows:

1. It can be observed that the compression times of different file types also show significant differences. For example, video files such as MKV files (54.733 ms) and MP4 files (38.026 ms) show longer compression times; while text files such as TXT files (0.302 ms) and XML files (1.129 s) show shorter compression times, due to the fact that video files usually have larger file sizes, while text files are usually smaller. As compression algorithms need to process more data blocks and complex data structures, the time required to process larger files is correspondingly longer. In contrast, smaller text files can be compressed more quickly due to their relatively simple structure.

2. Differences can be observed in the decompression times of different file types. Some file types, such as MKV files (6.656 ms), WAV files (5.777 ms), MP4 (3.997 ms) and other audio-video files, take relatively longer to decompress. This is due to the fact that audio-video files usually have larger file sizes and more complex decompression operations. Decompression involves a complex decoding process and the processing of multiple data channels, and therefore takes longer. In contrast, TXT files (0.067 ms) and XML files (0.19 ms) show shorter decompression times, which means that these files are relatively easy to decompress and have a faster decompression speed. Also, by comparing the compression time with the decompression time, it is seen that the decompression time is usually slightly shorter than the compression time. This is because the decompression operation does not require the computational process of the compression algorithm, but simply the reduction of the compressed data, and therefore is usually completed more quickly.

3. There are significant differences in compression rates for different file types. For example, people currently compress to varying degrees for common audio formats such as MP4 files (0.00%), WEBM files (0.12%) and MKV files (0.71%), but they still contain a large amount of repetitive and redundant information [7], which is not obvious when compressed again. As for the image file types, BMP image files exhibit a high compression rate of 82.40%, while JPG and PNG image files have lower compression rates of 0.66% and 0.28% respectively. This is due to the fact that BMP image files themselves do not use compression algorithms, while JPG and PNG image files use lossy and lossless compression algorithms, hence their lower compression ratios. In contrast, XML files (89.51%) and BIN files (86.93%) show higher compression ratios, indicating that these file types are able to significantly reduce their file size when compressed. The differences in compression rates reflect the data characteristics of the different file types and the suitability of the compression algorithms.

4. Anomalies were also found in the results. For the three file formats MKV, PNG and WEBM, some files were compressed to a larger file size than before. The MKV video file format often contains audio and video tracks that have already been compressed, and when the entire file is compressed again, the compression algorithm has difficulty providing additional compression and may even increase the file size slightly. The WEBM multimedia file format is commonly used for storing audio and video data and uses efficient audio codecs and video codecs, which may not provide significant additional compression when compressing the whole file again, or even slightly increase. In network transmission, for these file formats where the file size increases instead after compression and the compression ratio is very low, consider choosing the method of direct transmission without compression to minimise the time and resource consumption of data transmission. If the bandwidth is sufficient, the network is stable and the receiving end has sufficient processing power, then the choice of uncompressed transmission may also be a reasonable decision.

**4.2.2 BMP and TXT formats**

In the experiments, two file formats (BMP, TXT) that are significantly and commonly compressed were further selected for further comparative experiments. For each file format, seven datasets of different file sizes [13][15-16][21-30] were selected for decompression operations, and accurate records were kept to understand the file size, decompression time and compression ratio before and after compression, in order to fully evaluate the compression gains.

The size of the BMP dataset used in this experiment was approximately 13.58 GB and the size of the TXT dataset was approximately 0.23 GB. The results of the experimental data are shown in Table 3:

Table 3 Experimental data results

| Param⟍⟍Type | Number | Pre-compression size (MB) | Compressed size (MB) | Compression time (MS) | Decompression time (MS) | Compression rate (%) |
|---|---|---|---|---|---|---|
| BMP | 11686 | 0.000433 | 0.000128 | 0.000154 | 0.000043 | 70.44% |
| BMP | 6688 | 0.01086 | 0.008109 | 0.00098 | 0.000186 | 58.89% |
| BMP | 15557 | 0.019508 | 0.009805 | 0.001205 | 0.000188 | 60.97% |
| BMP | 4049 | 0.138848 | 0.102942 | 0.007094 | 0.001295 | 24.48% |
| BMP | 5513 | 0.148127 | 0.110608 | 0.006324 | 0.001205 | 25.33% |
| BMP | 15114 | 0.65723 | 0.060739 | 0.011658 | 0.003121 | 90.38% |
| BMP | 18971 | 0.750051 | 0.132008 | 0.022156 | 0.0036 | 82.40% |
| TXT | 4394 | 0.000151 | 0.000084 | 0.000184 | 0.000055 | 44.11% |
| TXT | 2584 | 0.001119 | 0.000565 | 0.000268 | 0.00006 | 49.16% |
| TXT | 7760 | 0.002006 | 0.001005 | 0.000302 | 0.000067 | 42.28% |
| TXT | 2000 | 0.003809 | 0.001796 | 0.000441 | 0.000092 | 50.57% |
| TXT | 19827 | 0.003972 | 0.001558 | 0.000397 | 0.000078 | 58.99% |
| TXT | 17561 | 0.006567 | 0.002865 | 0.000473 | 0.000092 | 42.27% |
| TXT | 1468 | 0.01357 | 0.004375 | 0.000879 | 0.000154 | 67.80% |

The compression time and decompression time for BMP format files and TXT format are shown in Figure 2:

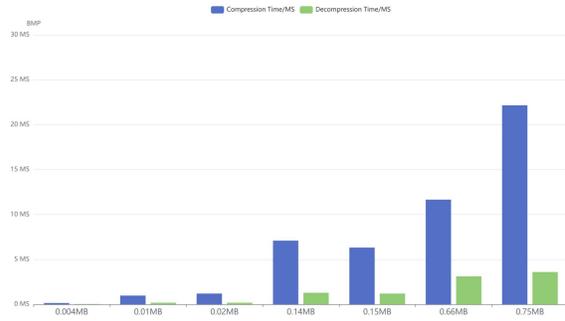

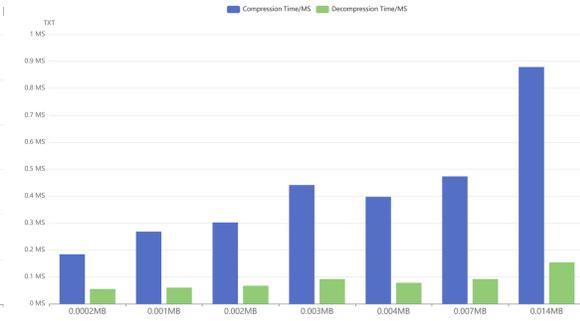

Figure 2.1                          Figure 2.2

Figure 2.1 shows the comparison of decompression time for BMP format files and Figure 2.2 shows the comparison of decompression time for TXT format files.

The compression ratio of files in BMP format and TXT format is shown in Figure 3:

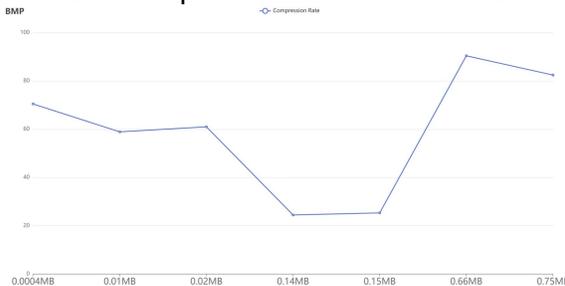

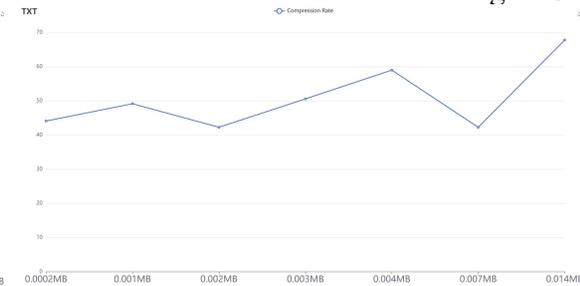

Figure 3.1                          Figure 3.2

Figure 3.1 shows the compression ratio for files in BMP format and Figure 3.2 shows the compression ratio for files in TXT format.

For the BMP image format and the TXT text format, the compression and decompression of files of different sizes can vary by orders of magnitude.

### 4.2.3 HDF5 and NetCDF formats

The massive data transfer also has practical significance for high performance computing, and two high performance computing file formats (HDF5 and NetCDF) were chosen for the compression experiments. The data set used in this experiment [39-53] is approximately 87 GB in size, of which the HDF5 file is approximately 51 GB and the NetCDF file is approximately 36 GB. Accurate records are kept to understand the file size, decompression time and compression ratio before and after compression in order to gain insight into the performance of these file formats in data transfer.

HDF5 (Hierarchical Data Format 5) is widely used in scientific research, data analysis, high performance computing and visualisation. An analysis of the decompression time and compression ratio of HDF5 format files is shown in Figure 4:

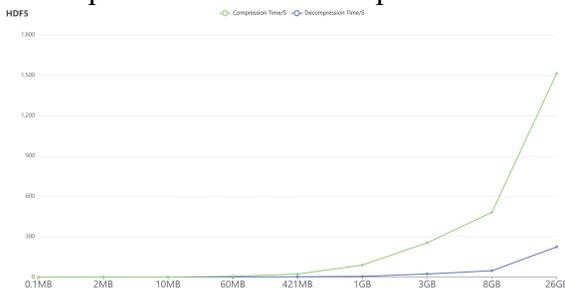

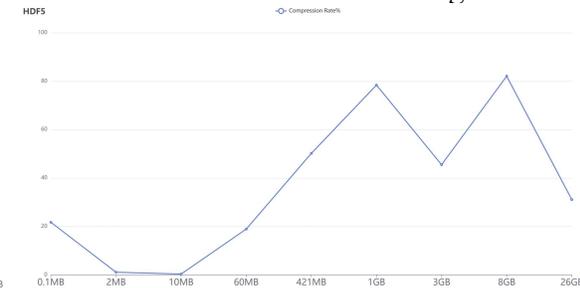

Figure 4.1                          Figure 4.2

Figure 4.1 shows the comparison of decompression time and compression ratio for HDF5 format files, and Figure 4.2 shows the compression ratio for HDF5 format files.

As the size of HDF5 format files increases, their decompression time shows a gradual increase. The decompression process for large files involves more demanding computational tasks and resource requirements, and as the file size increases, the computational resources required for the decompression operation also increases. Due to the need to process larger amounts of data, decompression algorithms must perform more computational operations during the execution phase, which inevitably leads to an increase in decompression time.

NetCDF (Network Common Data Format) is a file format for storing, accessing and sharing scientific data. An analysis of the decompression time and compression ratio of NetCDF format files is shown in Figure 5:

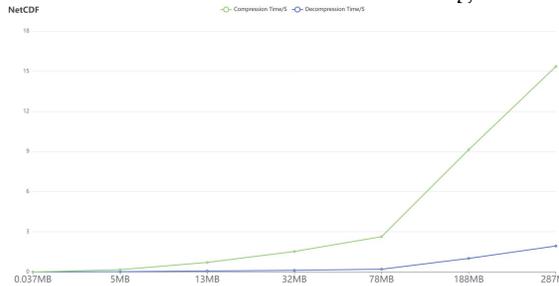
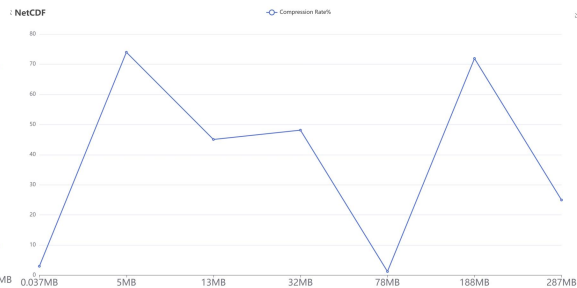

Figure 5.1       Figure 5.2

Figure 5.1 shows the comparison of decompression time and compression ratio of NetCDF format files, and Figure 5.2 shows the compression ratio of NetCDF format files.

The decompression time for NetCDF format files usually increases as the file size increases. The decompression process for large NetCDF files involves more disk read operations and computation operations, and larger files take longer to read data from the hard disk, which can increase the overall decompression time. In addition, the decompression algorithm may require more computation operations and memory resources to process more data, which can also lead to an increase in decompression time.

# 5. Conclusion

## 5.1 The conclusions of this experiment are as follows:

It was found through this experiment that there are significant differences in the compression performance of different types of file formats with different gains. Some file formats have higher compression ratios and significant compression effects, while others have lower compression effects. For file formats that have already been compressed (e.g. JPG, MP3, etc.), recompression may result in poor recompression due to the use of specific compression algorithms internally, which may result in information loss and repetitive compression. On the other hand, uncompressed file formats, such as text files and lossless image files (e.g. TXT, BMP), tend to exhibit high compression ratios and significant compression effects. This is because these file formats have a high level of redundancy and compressibility, and can be effectively reduced in file size by compression.

## 5.2 Future outlook:

This study focuses on the analysis of the compression performance of different file formats, but does not delve into the details of compression algorithms and optimisation methods. Future research could focus on improving and optimising existing compression algorithms to increase the compression and decompression speed while maintaining good compression quality.

# References

[1] Bai Winchao. Research and implementation of key algorithms for approximation calculation of massive data based on deep learning [D]. Harbin Institute of Technology. 2022.

[2] Qu Kaiyang. Design and implementation of contextual data compression algorithm based on improved Huffman [D]. Beijing University of Posts and Telecommunications,2017.

[3] Lu Bing,Liu Xinghai. Compression and decompression of files using improved Huffman coding[J]. Science and Technology Bulletin,2013,29(06):22-24.

[4] Mhd.Ali Subada.Comparisonal Analysis Of Even-Rodeh Algorithm Code AndFibonacci Code Algorithm For Text File Compression.Journal Basic Science and Technology.Feb 2022

[5] Wei Li. Integrity Control Techniques for Massive Data Transfer Under Big Data Cloud Storage [J]. Journal of Jilin University (Information Science Edition),2019,37(06):682-686.

[6] Zheng Dongxu. Research on transmission bandwidth optimization technology of satellite navigation system[D]. Shenyang University of Technology,2021.

[7] lzma — Compression using the LZMA algorithm — Python 3.11.3 documentation.

[8] Subin An.Kaggle Kerneler.（2021，May）Various Pokemon Image Dataset,Version 1,Retrieved May 25,2023 from Various Pokemon Image Dataset | Kaggle

[9] Theo Viel,Umong Sain,lnnat.(2022,December)RSNA Breast Cancer Detection - 512x512 pngs,Version 1,Retrieved May 26,2023 from RSNA Breast Cancer Detection - 512x512 pngs | Kaggle

[10] DeepMind.(2022,May)Kinetics dataset (5%),Version 1,Retrieved May 26,2023 from Kinetics dataset (5%) | Kaggle

[11] Michaël Defferrard, Kirell Benzi, Pierre Vandergheynst, Xavier Bresson, EPFL LTS2.(2017,May)FMA: A Dataset For Music Analysis Data Set,Version 1,Retrieved May 26,2023 from UCI Machine Learning Repository: FMA: A Dataset For Music Analysis Data Set.

[12] SyntheaTM.(2021,May)Synthea Dataset Jsons - EHR,Version 1,Retrieved May 26,2023 from Synthea Dataset Jsons - EHR | Kaggle

[13] Dipankar Srirag.Muhammad Navaid.Kaggle Kerneler.(2020,May)Text Classification on Emails，Version 1,Retrieved May 26,2023 from Text Classification on Emails | Kaggle

[14] Linjie Yang, Ping Luo, Chen Change Loy, Xiaoou Tang.A Large-Scale Car Dataset for Fine-Grained Categorization and Verification.Proceedings of the IEEE Conference on Computer Vision and Pattern Recognition (CVPR), 2015, pp. 3973-3981

[15] Md Fantacher lslamA.Matt OP.Guri.(2020,June)Metal Surface Defects Dataset,Version 1,Retrieved June 1,2023 from Metal Surface Defects Dataset | Kaggle.

[16] JinyeongWang.KaggleKerneler.Tannatorn Tantipiriyakul.(2019,June)Alphabet+Numbers.Version 1,Retrieved June 1,2023 from Alphabet+Numbers | Kaggle.

[17] Gzip - GNU Project - Free Software Foundation.

[18] bzip2 : Home (sourceware.org).

[19] https://github.com/madler/zlib

[20] KAIFENG YANG.(2022,June).Heat Sink Surface Defeect Dataset.Version 1,Retrieved June 2,2023 from Heat Sink Surface Defect Dataset | Kaggle.

[21] Gaurav Rajpal.Abdallah Wagih lbrahim.Beyza Nur Nakkas.(2020,June)Leukemia Classification,Version 1,Retrieved June 2,2023 from Leukemia Classification | Kaggle.

[22] Mohammed Aliy.Kaggle Kerneler.(2020,June)Single Cell Conventional Pap Smear Images,Version 1,Retrieved June 2,2023 from Single Cell Conventional Pap Smear Images | Kaggle.

[23] Dustin Ober.(2023,April)Soybean Seeds Classification Dataset,Version 1,Retrieved June 2,2023 from Soybean Seeds| Kaggle.

[24] John Margaronis.Minas Christou.Ergina Kavallieratou.Theodoros Tzouramanis.(2020,June)Handwritten Greek Characters from GCDB,Version 1,Retrieved June 2,2023 from Handwritten Greek Characters from GCDB | Kaggle.

[25] RNA.Kaggle Kerneler.(2020,June)SocioEconomic Data and Applications Center,Version 1,Retrieved June 3,2023 from SocioEconomic Data and Applications Center | Kaggle.

[26] Kaggle Kerneler.Noah Daniels.(2019,June)Sports articles for objectivity analysis,Version 1,Retrieved June 3,2023 from Sports articles for objectivity analysis | Kaggle.

[27] kobeshigaidaicorpus.Sami Hirata.Kaggle Kerneler.(2020,June)NICT Japanese Learners of English 4.1,Version 1,Retrieved June 3,2023 from NICT Japanese Learners of English 4.1 | Kaggle.

[28] Hsankesara.Paul Mooney.stpete_ishii.(2019,June)CVPR 2019 Papers,Version 1,Retrieved June 3,2023 from CVPR 2019 Papers | Kaggle.

[29] MihxSP.Kaggle Kerneler.Mike Klemin(2020,June)Kinopoisk's movies reviews,Version 1,Retrieved June 3,2023 from Kinopoisk's movies reviews | Kaggle.

[30] Alien.Huan YEH.Kaggle Kerneler.(2021,June)vinbigdata txt yolov5Version 1,Retrieved June 3,2023 from vinbigdata txt yolov5 | Kaggle.

[31] Liu Xiaoyan, Xu Zhiwei, Li Wenyue et al. Classifiable data compression transmission mechanism for efficient edge computing[J]. Software Guide,2022,21(11):38-43.

[32] Li M, Yin S, Zhang N, et al. Sparse representation-based lossless compressed transmission method for array acoustic logging data[J]. Measurement and Control Technology,2022,41(05):106-112.

[33] Wang Julong, Zhang Huaizhu, Xuan Jinpeng. Research on data acquisition and compressed transmission of broadband seismometers [U]. Modern Electronics Technology,2020,43(04):100-103.

[34] Yan L, Li YB. Research on methods of effective data compression in computer network transmission [J]. Communication World,2016,(15):20-21.

[35] Yang Jingfeng,Zhang Nanfeng,Li Yong et al. Compression method for data transmission of agricultural machinery operations based on improved Huffman coding[J]. Journal of Agricultural Engineering,2014,30(13):153-159.

[36] Peng Chong. Research and application of data compression and transmission optimization algorithm for sensor network big data transmission application[D]. University of Electronic Science and Technology, 2014.

[37] Ma Xingming,Dong Cheng,Mao Xinyu et al. State estimation-based data compression and storage method for massive multivariate heterogeneous smart grid[J]. Motor and Control Applications,2023,50(02):67-72+81.

[38] Wang H, Li Shiqiang, Yu Huainan et al. A compressed storage method for power quality data in distribution networks based on distributed compressed sensing and edge computing[J]. Journal of Electrical Engineering Technology,2020,35(21):4553-4564.

[39] SUNGHO SHIM.（2022，Augest）Tiny ImageNet (HDF5),Version 1,Retrieved June 7,2023 from Tiny ImageNet (HDF5) | Kaggle.

[40] Kaggle Kerneler.Benedict Wilkins.Limon Halder.(2020,June).MNIST - HDF5,Version 1,Retrieved June 7,2023 from MNIST - HDF5 | Kaggle.

[41] MUHAMMAD IRFAN AZAM.(2022,June).AMEX HDF5 - Last Statement - Train Only,Version 1,Retrieved June 7,2023 from AMEX HDF5 - Last Statement - Train Only | Kaggle.

[42] Kagle Kerneler.Olga Belitskaya.(2020,June).Quick, Draw! Model Weights for Doodle Recognition,Version 1,Retrieved June 8,2023 from Quick, Draw! Model Weights for Doodle Recognition | Kaggle.

[43] Valentyn Sichkar.(2021,June).Traffic Signs 1 million images for Classification,Version 1,Retrieved June 8,2023 from Traffic Signs 1 million images for Classification | Kaggle.

[44] K Scott Mader. 孙 健 行 James_97_soton.Wouter van Amsterdam.(2017,June).Lung Nodule Malignancy,Version 1,Retrieved June 8,2023 from Lung Nodule Malignancy | Kaggle.

[45] ladyofateele.Habineza.Elena Cuoco.(2019,June).The Gravitational Waves Discovery Data, Version 1,Retrieved June 8,2023 from The Gravitational Waves Discovery Data | Kaggle.

[46] Marco Polo.Richard Kuo.Asmaa.(2020,June).Brain Tumor Segmentation(BraTS2020),Version 1,Retrieved June 8,2023 from Brain Tumor Segmentation(BraTS2020) | Kaggle.

[47] MC1138.（2022,October）.GISTEMP seasonal trends,Version 1,Retrieved June 8,2023 from GISTEMP seasonal trends | Kaggle.

[48] Baris Dincer.(2021,June).CLIMATE CHANGE MADAGASCAR-TURKEY /NASA,Version 1,Retrieved June 8,2023 from CLIMATE CHANGE MADAGASCAR - TURKEY / NASA | Kaggle.

[49] Gabriel Preda.Baris Dincer.Kaggle Kerneler.(2020,June).EarthData MERRA2 Co,Version 1,Retrieved June 8,2023 from [EarthData MERRA2 CO | Kaggle](#).

[50] Paula Romero Jure.Kaggle Kerneler(2020,June).GOES_L1,Version 1,Retrieved June 8,2023 from [GOES_L1 | Kaggle](#).

[51] Abhinav Sharma.(2021,June).ERA data files - daily(122) for selected domain,Version 1,Retrieved June 8,2023 from [ERA data files - daily(122) for selected domain | Kaggle](#).

[52] Vincentive Larmet.Kaggle Kerneler.(2020,June).ERA interim wind data in Puerto Rico,Version 1,Retrieved June 8,2023 from [ERA interim wind data in Puerto Rico | Kaggle](#).

[53] Parichat Wetchayont.(2021,January).OSTIA SST Asia,Version 1,Retrieved June 8,2023 from [OSTIA SST Asia | Kaggle](#).

# Figure Legends

[Insert Figure Legends here]

要求：将文中所有图题和图例放在此页。按照在正文中引用的顺序排列。

Fig. 1

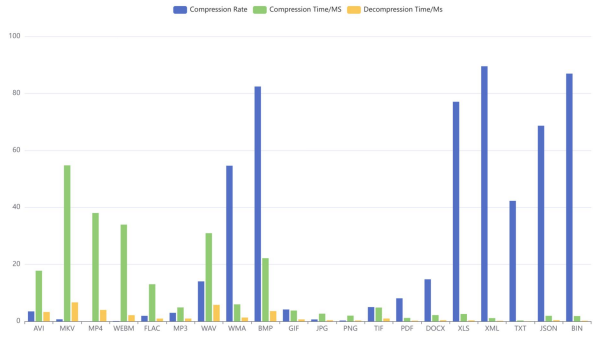

Fig. 2

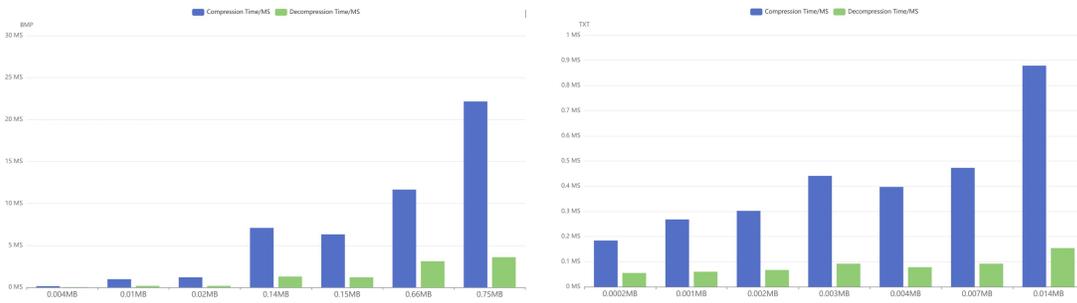

Fig. 3

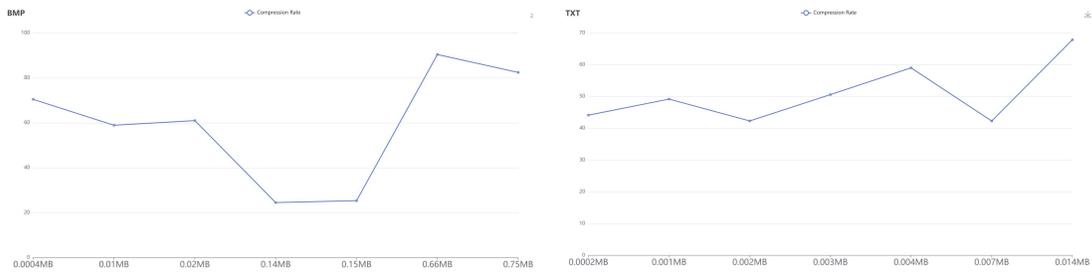

Fig. 4

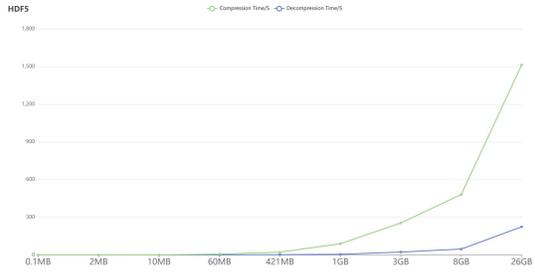

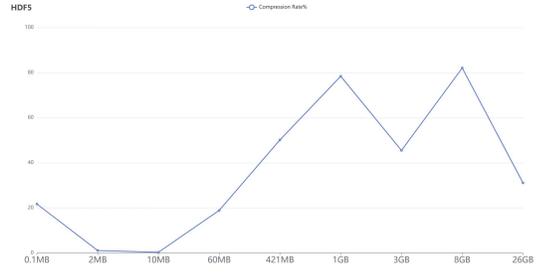

Fig. 5

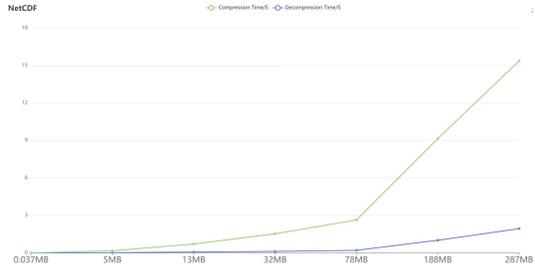

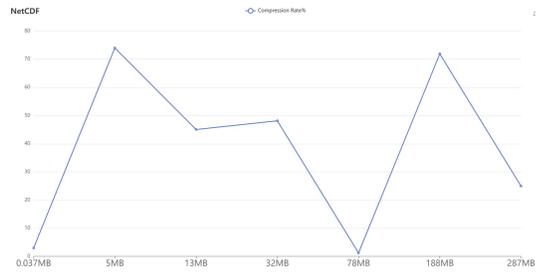

# Figures

[Insert Figures here.]

格式要求：
图字、表字；图题、表题；图注和表注：TNR，8 pt。
图表尺寸：单栏最大：185 mm × 120 mm；双栏最大：宽不超过 60 mm。
图清晰度：300 dpi 或以上。若图中有多个图表，用 A、B、C 等分别表示。
注明图（横纵坐标）表的度量衡单位及其说明。
按照正文中引用的顺序升序排列；每个图单独占据一页，不同图之间用分页符。

公式：
公式放在正文中，需用公式编辑器进行编辑。

**Tables**

Table 1 Experimental data set file

| Dataset name | File formats | File type |
|---|---|---|
| Transferred from Kinetics dataset | Video | AVI |
| Transferred from Kinetics dataset | Video | MKV |
| Kinetics dataset[10] | Video | MP4 |
| Transferred from Kinetics dataset | Video | WEBM |
| Transferred from FMA | Audio | FLAC |
| FMA: A Dataset For Music Analysis Data Set[11] | Audio | MP3 |
| Transferred from FMA | Audio | WAV |
| Transferred from FMA | Audio | WMA |
| Metal Surface Defects Dataset[15] | Image | BMP |
| Synthea Dataset Jsons - EHR[12] | Image | GIF |
| A Large-Scale Car Dataset for Fine-Grained Categorization and Verification. (CVPR)[14] | Image | JPG |
| RSNA Breast Cancer Detection - 512x512 pngs[9] | Image | PNG |
| Transferred from A Large-Scale Car Dataset for Fine | Image | TIF |
| Transferred from DOCX | Documentation | PDF |
| Self-collection | Documentation | DOCX |
| Self-collection | Documentation | XLS |
| Self-collection | Text | XML |
| Text Classification on Emails[13] | Text | TXT |
| Various Pokemon Image Dataset[8] | | JSON |
| Self-collection | Binary | BIN |

Table 2 Experimental data results

| Param \ Type | Number | Pre-compression size (MB) | Compressed size (MB) | Compression time (MS) | Decompression time (MS) | Compression rate (%) |
|---|---|---|---|---|---|---|
| AVI | 12748 | 0.522941 | 0.508537 | 17.76 | 3.28 | 3.49% |
| MKV | 10350 | 2.105637 | 2.102279 | 54.733 | 6.656 | 0.71% |
| MP4 | 10214 | 1.469854 | 1.463328 | 38.026 | 3.997 | 0.00 |
| WEBM | 6653 | 1.200335 | 1.20309 | 33.938 | 2.162 | 0.12% |
| FLAC | 8733 | 0.581497 | 0.574143 | 13.019 | 0.963 | 1.93% |
| MP3 | 8732 | 0.13802 | 0.134574 | 4.877 | 0.972 | 2.98% |
| WAV | 8733 | 0.774194 | 0.660883 | 30.932 | 5.777 | 14.02% |
| WMA | 8733 | 0.242912 | 0.109045 | 5.96 | 1.326 | 54.61% |
| BMP | 15557 | 0.750051 | 0.132008 | 22.156 | 3.6 | 82.40% |
| GIF | 7261 | 0.08965 | 0.086222 | 3.781 | 0.678 | 4.16% |
| JPG | 10547 | 0.089225 | 0.088777 | 2.693 | 0.392 | 0.66% |
| PNG | 54707 | 0.065476 | 0.065365 | 1.996 | 0.287 | 0.28% |
| TIF | 8895 | 0.140895 | 0.135055 | 4.835 | 0.981 | 5.00% |
| PDF | 2485 | 0.02381 | 0.021898 | 1.193 | 0.225 | 8.08% |
| DOCX | 2485 | 0.051761 | 0.043766 | 2.193 | 0.389 | 14.75% |
| XLS | 15121 | 0.052615 | 0.012066 | 2.582 | 0.324 | 77.05% |
| XML | 3912 | 0.043191 | 0.003706 | 1.129 | 0.19 | 89.51% |
| TXT | 7760 | 0.002006 | 0.001005 | 0.302 | 0.67 | 42.28% |
| JSON | 10628 | 0.080233 | 0.019754 | 1.926 | 0.399 | 68.64% |
| BIN | 1500 | 0.029862 | 0.003754 | 1.861 | 0.198 | 86.93% |

Table 3 Experimental data results

| Param \ Type | Number | Pre-compresion size (MB) | Compressed size (MB) | Compression time (MS) | Decompression time (MS) | Compression rate（%） |
|---|---|---|---|---|---|---|
| BMP | 11686 | 0.000433 | 0.000128 | 0.000154 | 0.000043 | 70.44% |
| BMP | 6688 | 0.01086 | 0.008109 | 0.00098 | 0.000186 | 58.89% |
| BMP | 15557 | 0.019508 | 0.009805 | 0.001205 | 0.000188 | 60.97% |
| BMP | 4049 | 0.138848 | 0.102942 | 0.007094 | 0.001295 | 24.48% |
| BMP | 5513 | 0.148127 | 0.110608 | 0.006324 | 0.001205 | 25.33% |
| BMP | 15114 | 0.65723 | 0.060739 | 0.011658 | 0.003121 | 90.38% |
| BMP | 18971 | 0.750051 | 0.132008 | 0.022156 | 0.0036 | 82.40% |
| TXT | 4394 | 0.000151 | 0.000084 | 0.000184 | 0.000055 | 44.11% |
| TXT | 2584 | 0.001119 | 0.000565 | 0.000268 | 0.00006 | 49.16% |
| TXT | 7760 | 0.002006 | 0.001005 | 0.000302 | 0.000067 | 42.28% |
| TXT | 2000 | 0.003809 | 0.001796 | 0.000441 | 0.000092 | 50.57% |
| TXT | 19827 | 0.003972 | 0.001558 | 0.000397 | 0.000078 | 58.99% |
| TXT | 17561 | 0.006567 | 0.002865 | 0.000473 | 0.000092 | 42.27% |
| TXT | 1468 | 0.01357 | 0.004375 | 0.000879 | 0.000154 | 67.80% |

要求：表头、表题和表均放置在此处。不同表占据不同页码，表之间用分页符分开。
采用三线表。要求简洁、自明性强。
图字、表字；图题、表题；图注和表注：TNR，8 pt。
图表尺寸：单栏最大：185 mm × 120 mm；双栏最大：宽不超过 60 mm。